\documentclass[aps,nofootinbib,twocolumn,prd,eqsecnum,showpacs,showkeys,preprintnumbers]{revtex4-1}
\usepackage{graphicx}
\usepackage{setspace}
\usepackage{adjustbox}
\usepackage{amsmath}
\usepackage{amsfonts}
\usepackage{amssymb}
\usepackage{color}
\usepackage{xcolor}
\usepackage{balance}
\usepackage{bm}
\usepackage{float}
\usepackage{mathrsfs}
\usepackage{epstopdf}
\usepackage{url}
\usepackage{booktabs}
\usepackage{footnote}
\usepackage{textcomp}
\usepackage[normalem]{ulem}
\usepackage{esint}
\usepackage[unicode=true, pdfusetitle,
 bookmarks=true,bookmarksnumbered=false,bookmarksopen=false,
 breaklinks=false,pdfborder={0 0 1},backref=false,colorlinks=false]{hyperref}
\usepackage{multirow}
\usepackage{pifont}
\usepackage{times}
\usepackage[english]{babel}

\usepackage{appendix}
\usepackage{float}
\usepackage{enumerate}
\usepackage{lineno}

\usepackage{hyperref}

\usepackage{tabularx}

\makeatletter

\newcommand{\stkout}[1]{\ifmmode\text{\sout{\ensuremath{#1}}}\else\sout{#1}\fi}

\AtBeginDocument{
\heavyrulewidth=.08em
\lightrulewidth=.05em
\cmidrulewidth=.03em
\belowrulesep=.65ex
\belowbottomsep=0pt
\aboverulesep=.4ex
\abovetopsep=0pt
\cmidrulesep=\doublerulesep
\cmidrulekern=.5em
\defaultaddspace=.5em
}

\newcolumntype{L}[1]{>{\hsize=#1\hsize\raggedright\arraybackslash}X}%
\newcolumntype{R}[1]{>{\hsize=#1\hsize\raggedleft\arraybackslash}X}%
\newcolumntype{C}[1]{>{\hsize=#1\hsize\centering\arraybackslash}X}%

\newcommand*\patchAmsMathEnvironmentForLineno[1]{%
 \expandafter\let\csname old#1\expandafter\endcsname\csname #1\endcsname
 \expandafter\let\csname oldend#1\expandafter\endcsname\csname end#1\endcsname
 \renewenvironment{#1}%
   {\linenomath\csname old#1\endcsname}%
   {\csname oldend#1\endcsname\endlinenomath}}%
\newcommand*\patchBothAmsMathEnvironmentsForLineno[1]{%
 \patchAmsMathEnvironmentForLineno{#1}%
 \patchAmsMathEnvironmentForLineno{#1*}}%
\AtBeginDocument{%
\patchBothAmsMathEnvironmentsForLineno{equation}%
\patchBothAmsMathEnvironmentsForLineno{align}%
\patchBothAmsMathEnvironmentsForLineno{flalign}%
\patchBothAmsMathEnvironmentsForLineno{alignat}%
\patchBothAmsMathEnvironmentsForLineno{gather}%
\patchBothAmsMathEnvironmentsForLineno{multline}%
}

\begin{document}

\title{Cosmological constraints of interacting phantom dark energy models}

\author{Amine Bouali$^{1}$}
\email{a1.bouali@upm.ac.ma}
\author{Imanol Albarran $^{2,3}$}
\email{imanolalbarran@gmail.com}
\author{Mariam Bouhmadi-L\'{o}pez$^{4,5}$}
\email{{\mbox{mariam.bouhmadi@ehu.eus}}}
\author{Ahmed Errahmani$^{1}$}
\email{ahmederrahmani1@yahoo.fr}
\author{Taoufik Ouali$^{1}$}
\email{ouali1962@gmail.com}

\date{\today }
\affiliation{
${}^1$Laboratory of Physics of Matter and Radiation, Mohammed I University, BP 717, Oujda, Morocco\\
${}^2$Departamento de F\'{\i}sica, Universidade da Beira Interior, Rua Marqu\^{e}s D'\'Avila e Bolama 6200-001 Covilh\~a, Portugal\\
${}^3$Centro de Matem\'atica e Aplica\c{c}\~oes da Universidade da Beira Interior, Rua Marqu\^{e}s D'\'Avila e Bolama 6200-001 Covilh\~a, Portugal\\
${}^4$IKERBASQUE, Basque Foundation for Science, 48011, Bilbao, Spain\\
${}^5$Department of Physics University of the Basque Country UPV/EHU. P.O. Box 644, 48080 Bilbao, Spain\\
}

\begin{abstract}
In this paper, we consider three phantom dark energy models, in the context of interaction between the dark components namely cold dark matter (CDM) and dark energy (DE). The first model, known as $w_{\textrm{d}}$CDM can  induce a big rip singularity (BR) while the two remaining induce future abrupt events known as the Little Rip (LR) and Little Sibling of the Big Rip (LSBR). These phantom DE models can be distinguished by their equation of state. 
We invoke a new phenomenon such as the interaction between CDM and DE given that it could solve or alleviate some of the problems encountered in standard cosmology. We aim to find out the effect of such an interaction on the cosmological parameters of the studied  models,  as well as, the persistence or the disappearance of the singularity and the abrupt events induced by  the  models under study. We choose an interaction term proportional to DE density, i.e. $Q=\lambda H \rho_{\textrm{d}}$, since the case where $Q\propto \rho_{\textrm{m}}$ could lead to a large scale instability at  early time. We also do not claim at all that $Q=\lambda H \rho_{\textrm{d}}$ is the ideal choice since it suffers from a negative CDM  density in the future. By the use of a Markov Chain Monte Carlo (MCMC) approach, and by assuming  a flat FLRW Universe, we constrain the cosmological parameters of each of the three phantom DE models studied. Furthermore, by the aid of  the corrected Akaike Information Criterion ($\text{AIC}_{c}$) tool, we  compare our phantom DE models. Finally, a perturbative analysis of phantom DE models under consideration is performed based on the  best fit background parameters.
\end{abstract}

\maketitle

\section{INTRODUCTION}\label{sec1}
The observations of distant type Ia supernovae (SNIa), the Cosmic Microwave Background measurements (CMB) and the Baryon acoustic Oscillations (BAO) \cite{Riess:1998cb, Perlmutter:1998np, Ade:2015rim} indicate that the expansion of the Universe is presently accelerating. The origin behind this acceleration is still  a mystery up to date.  One of the most accepted explanations to this phenomenon is the suggestion that the Universe is filled with  a  dark component known as dark energy (DE). The most popular model up to now to explain accurately the data is $\Lambda$CDM, where $\Lambda$ is the cosmological constant and it plays the role of dark energy. Even though, $\Lambda$CDM  explains the observational data almost on all scales, it suffers from many problems, such as the  coincidence problem \cite{Cai:2004dk}, the age problem caused by old quasars \cite{Duran:2010ky}. Over the years, while researchers suggested many models that are basically extensions of $\Lambda$CDM model in the context of general relativity for different types of scalar fields, like quintessence \cite{Ratra:1987rm}, k-essence \cite{ArmendarizPicon:2000ah}  or for different  kind of equations of state such as the  Chaplygin gas \cite{Kamenshchik:2001cp}, other extensions based on  modified theories of gravity like f(R)-gravity \cite{Sotiriou:2008rp,DeFelice:2010aj}, f(T)-gravity \cite{Ferraro:2006jd,Ferraro:2008ey,Bengochea:2008gz} and f(R,T)-gravity \cite{Harko:2011kv,Nojiri:2004bi,Allemandi:2005qs,Nojiri:2006gh,Nojiri:2009kx} have been also proposed as an alternative mechanism to describe the current acceleration of the Universe.  In this paper, we will focus on the first aforementioned class of DE.

Our knowledge about the dark components and their features is still very limited, since they are only detected indirectly via their gravitational effects, which makes their nature very secretive. One of the innovative ideas in modern cosmology is to consider an interaction between the dark components i.e. CDM and DE, which  allows an energy transfer between them. Suggesting DE models that include new effects like  interaction between DE and CDM will provide not only the possibility to investigate the dark sector on a completely new level, but also could help us to discriminate between different theoretical models. Furthermore, such interactions could help to solve or alleviate the standard cosmology problems cited above.

The discovery that our Universe is accelerating has motivated theorists to infer the best theoretical model from observations. As a result,  hundreds of theoretical models have been proposed to this aim based on different theories. Some of these models present singularities i.e. cosmological quantities that go to infinity at a finite cosmic time, while others present abrupt events, i.e. cosmological quantities that diverge at an infinite cosmic time. The equation of state (EoS) of DE models which is the ratio between the pressure and the energy density can lead to different fates of the Universe. In general, we can classify DE models in two classes: the first class corresponds to an EoS larger than -1, this type of models is known as quintessence, the second class corresponds to models having  an EoS smaller than -1 and are known as phantom models. Surprisingly, and even though they violate the null energy condition, these models are the most supported by cosmological observations \cite{Aghanim:2018eyx,Caldwell:1999ew,Jimenez:2016sgs,Sahni:2014ooa,Vagnozzi:2018jhn,Alam:2016wpf} rather than being ruled out as it was expected.
In the current paper, we will focus on three phantom DE models, that induce different doomsday known as Big Rip (BR) \cite{Caldwell:1999ew,Dabrowski:2003jm,Starobinsky:1999yw,Caldwell:2003vq,Carroll:2003st,Chimento:2003qy,GonzalezDiaz:2003rf,GonzalezDiaz:2004vq,Sahni:2002dx} Little Rip (LR) \cite{Ruzmaikina,Bouhmadi-Lopez:2013nma,Nojiri:2005sx,Nojiri:2005sr,Stefancic:2004kb,BouhmadiLopez:2005gk,Frampton:2011sp,Brevik:2011mm,Contreras:2018two} and Little Sibling of the Big Rip (LSBR) \cite{Bouhmadi-Lopez:2014cca,Morais:2016bev,Morais:2016bev,Bouhmadi-Lopez:2018lly}. \par
In absence of interaction between DE and CDM the BR model induces a true singularity while the two remnants induce a future abrupt events. In a previous work \cite{Bouali:2019whr,Albarran:2016mdu}, we have studied the  same set of models without interaction between DE and CDM components, the analysis was done at the background and the perturbative levels by means of a  Markov Chain Monte Carlo (MCMC) at the background level \cite{Sagredo:2018ahx}. In the current paper, we study the same  models by taking into account an interaction between DE and CDM in order to find out  how the exchange of energy density between the dark components could modify  the  models  suitability according to observations. Furthermore, it may enhances the statistical evidence of the  models, in particular, as compared to $\Lambda$CDM. In addition, general relativity could  describe the background dynamics of the Universe even when such kind of interactions is taken into account. The  interaction  between CDM and DE densities is usually introduced phenomenologically given the absence of a theory determining the choice of  such form. Among the years many forms have been tested such as $Q=3\lambda H \rho_{\textrm{m}}$, $Q=3\lambda H \rho_{\textrm{d}}$ and $Q=3\lambda H \frac{\rho_{\textrm{d}}\rho_{\textrm{m}}}{\rho_{\textrm{d}}+\rho_{\textrm{m}}}$ among others \cite{Amendola:1999qq,Billyard:2000bh,Zimdahl:2001ar,Farrar:2003uw,Chimento:2003iea,Olivares:2005tb,Koivisto:2005nr,Sadjadi:2006qp,Guo:2007zk,Zhang:2007uh,Boehmer:2008av,Pereira:2008at,He:2010ta,Li:2011ga,Clemson:2011an,Xu:2013jma,Cheng:2019bkh,epjc2018,ijmp2020}. 
In this paper, we consider the following interaction $Q=\lambda H \rho_{\textrm{d}}$, where a positive coupling $\lambda >0$  ensures an energy transfer from CDM to DE density, while a negative coupling $\lambda <0$ ensures an energy transfer from DE to CDM density. Thus, the continuity equations are  written as follows:
\begin{equation}\label{coupled}
\left\{
\begin{array}{l}
 \dot{\rho}_{\textrm{m}}+3H\rho_{\textrm{m}}=-Q,\\
\dot{\rho}_{\textrm{d}}+3H(1+w_{\textrm{d}})\rho_{\textrm{d}}=Q,
\end{array}
\right.
\end{equation} 
where $\rho_{\textrm{m}}$, $\rho_{\textrm{d}}$ and $\lambda$ are the CDM, DE energies densities and the strength of the interaction, respectively, where for simplicity we choose $Q=\lambda H\rho_{\textrm{d}}$. 
The standard form of the dimensionless Friedmann equation is
\begin{equation}\label{FE}
E(a)^2=\Omega_{\textrm{r}}+\Omega_{\textrm{m}}+\Omega_{\textrm{d}},
\end{equation}
where $E(a)=H/H_0$ and
\begin{equation}\label{defOmega}
\Omega_{\textrm{r}}=\frac{8\pi G}{3H_0^2}\rho_{\textrm{r}}, \;\Omega_{\textrm{m}}=\frac{8\pi G}{3H_0^2}\rho_{\textrm{m}}\;\text{and}\;\Omega_{\textrm{d}}=\frac{8\pi G}{3H_0^2}\rho_{\textrm{d}},
\end{equation}
are the fractional energy densities of radiation, CDM and DE, respectively.  From now on, the subindex $0$ will stand for values fixed at present, therefore,  $\Omega_{\textrm{r0}}+\Omega_{\textrm{m0}}+\Omega_{\textrm{d0}}=1$ is satisfied.\par

\

The combination of  observational probes we use to constraint our phantom DE models are the Pantheon compilation of SNIa dataset \cite{Scolnic:2017caz}, Planck 2018 distance priors of CMB \cite{Zhai:2018vmm,Aghanim:2018eyx}, BAO measurements including 6dFGS, SDSS DR7 MGS, BOSS-LOWZ, BOSS-DR12, WiggleZ, BOSS-CMASS, Lya and DES \cite{Anderson:2013zyy,Beutler:2011hx,Ross:2014qpa,Kazin:2014qga,Alam:2016hwk} and the direct measurements of the Hubble parameter \cite{Anderson:2013zyy,Zhang:2012mp,Stern:2009ep,Moresco:2012jh,Chuang:2012qt,Moresco:2015cya,Moresco:2016mzx,Stocker:2018avm}. After extracting the best fit parameters and their corresponding minimal $\chi^{2}$ values, we classify our models according to their ability to fit the observational data by using the so called  Akaike Information Criterion ($\text{AIC}_{c}$), the results of this analysis are summarised in table \ref{tab1}.

\

\begin{figure}
\centering
\includegraphics[scale=0.55]{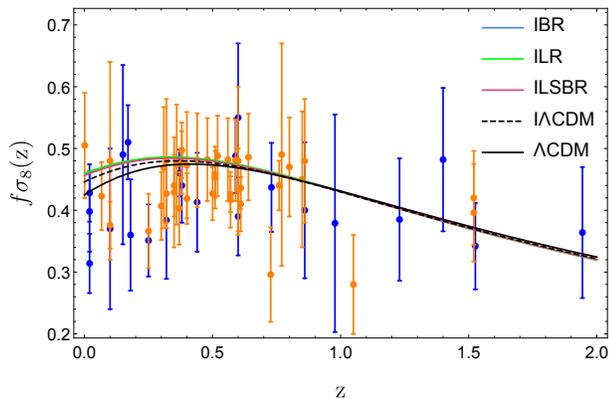} 
\caption{  This plot shows the evolution of the predicted $f \sigma_8$ of IBR, ILR, ILSBR and $\Lambda$CDM model against the observational growth rate data shown in table \ref{tab4} . The black curve corresponds to the $\Lambda$CDM, while the light blue, the green and the pink curves correspond to IBR, ILR and ILSBR models, respectively. The blue points correspond to RSD-22 compilation while the remaining points of RSD-63 are plotted in orange color.}\label{fs8}
\end{figure}

\begin{figure}
\centering
\includegraphics[scale=0.55]{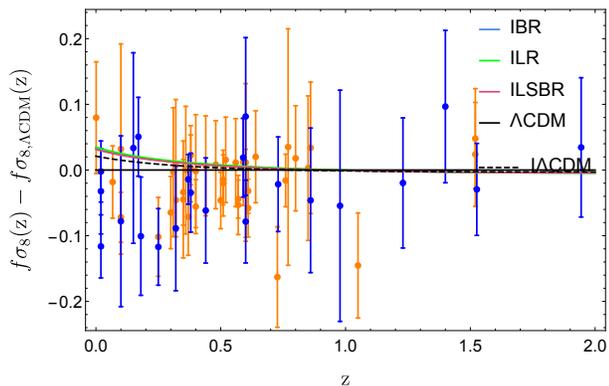} 
\caption{  This plot shows the difference of the predicted $f \sigma_8$ of IBR, ILR and ILSBR model with respect to that of the $\Lambda$CDM. The black curve corresponds to the $\Lambda$CDM, while the light blue, the green and the pink curve correspond to IBR, ILR and ILSBR models, respectively. The blue points correspond to RSD-22 compilation while the remaining points of RSD-63 are plotted in orange color.}\label{delta_fs8}
\end{figure}

The paper is structured as follows:  in Section~\ref{notes} we review the induced singularities and abrupt events without interaction for a generalised EoS which unifies the models addressed in the present work. In section~\ref{sec2} we give a short description of the three phantom DE models studied in this work and the observational constraints through the MCMC confidence contours for each of them. In section \ref{induced_events}, we analyse the asymptotic behaviour of $\dot{H}$ in order to classify the induced future abrupt event when considering interaction between DE and CDM. In section \ref{sec3},  we present the MCMC results in the table \ref{tab1},  we show also the evolution of the EoS of these three models  with respect to that of $\Lambda$CDM. Section \ref{sec4} is dedicated to the first order perturbation analysis where the results are shown in table \ref{tab3}.   We show also in the same section (see figures \ref{fs8} and \ref{delta_fs8}) the theoretical predictions of $f\sigma_8$ for each models,  and compare them with those corresponding to  $\Lambda$CDM. Finally, section \ref{sec5} is dedicated to  the conclusions.

\section{Non-interacting DE}\label{notes}

In this section we analyse a generalised EoS that unifies the DE models considered in the present work, these paradigms, obey the following EoS:
\begin{itemize}
\item Model A: $p_{\textrm{d}}=w_{\textrm{d}}\rho_{\textrm{d}}$.
\item Model B: $p_{\textrm{d}}=-\left( \rho_{\textrm{d}}+B\rho_{\textrm{d}}^{\frac{1}{2}}\right)$.
\item Model C: $p_{\textrm{d}}=-\left( \rho_{\textrm{d}}+\frac{C}{3}\right).$
\end{itemize}

Therefore, we could write a generalised EoS that gathers the above mentioned models in a single equation:
\begin{equation}\label{genEoS}
p_{\textrm{d}}=-\left( \rho_{\textrm{d}}+ A\rho_{\textrm{d}}^{\alpha}\right),
\end{equation}
where the constants $A$ and $\alpha$ characterise the model. We assume a positive value for $A$ for the fluid to be a phantom type of DE. Our models A, B and C correspond to, $A=-(w_{\textrm{d}}+1),B,C/3$ and $\alpha=1,1/2,0$  respectively. We will consider a single DE component. Using the conservation equation $\dot{\rho_{\textrm{d}}}+3H\left(p_{\textrm{d}}+\rho_{\textrm{d}}\right)=0$, where dot stands for derivative with respect the cosmic time,  we get:
\begin{equation}\label{genrhodif}
 \frac{d\rho_{\textrm{d}}}{da}=\frac{3A}{a}\rho_{\textrm{d}}^{\alpha}.
\end{equation}
Solving the equation (\ref{genrhodif}) and choosing $\rho(a_{0})=\rho_{0}$, where $a_{0}=1$, we get\footnote{The case for $\alpha=1$ is a special case, however, its solution is the well known DE  model  for a constant EoS parameter, i.e. $\rho_{\textrm{d}}=\rho_{\textrm{d}0} a^{3A}$. In addition, a subscript $0$ stands for quantities evaluated at present.}
\begin{equation}\label{genrho}
\rho_{\textrm{d}}=\left[3A\left(1-\alpha\right)x+\rho_{\textrm{d}0}^{1-\alpha}\right]^\frac{1}{1-\alpha},
\end{equation}
where we have defined $x\equiv\ln{(a)}$, therefore, $\dot{x}\equiv H$ and derivatives over time transform as $\dot{\left\{\phantom{a}\right\}}=H{\left\{\phantom{a}\right\}}_{x}$, where the subindex $x$ stands for derivatives over  $x$. Therefore, the Hubble parameter and its cosmic time derivatives could be written as:
\begin{align}
H&=\left(\frac{8\pi G}{3}\right)^{\frac{1}{2}}\left[3A\left(1-\alpha\right)x+\rho_{\textrm{d}0}^{1-\alpha}\right]^\frac{1}{2\left(1-\alpha\right)} ,\label{genH}
 \\
\dot{H}&=\left(\frac{8\pi G}{3}\right)\left(\frac{3A}{2}\right)\left[3A\left(1-\alpha\right)x+\rho_{\textrm{d}0}^{1-\alpha}\right]^\frac{2\alpha}{2\left(1-\alpha\right)},\label{genH1}
 \\
\ddot{H}&=\left(\frac{8\pi G}{3}\right)^{\frac{3}{2}}\left(\frac{3A}{2}\right)^{2}2\alpha\left[3A\left(1-\alpha\right)x+\rho_{\textrm{d}0}^{1-\alpha}\right]^\frac{4\alpha-1}{2\left(1-\alpha\right)},\label{genH2} \\
&\vdots\nonumber 
\\
H^{(n)}&=\left(\frac{8\pi G}{3}\right)^{\frac{n+1}{2}}\left(\frac{3A}{2}\right)^{n}\prod^{n-1}_{j=0}\left(2\alpha j -j+1\right)\nonumber
\\
&\times\left[3A\left(1-\alpha\right)x+\rho_{\textrm{d}0}^{1-\alpha}\right]^\frac{n\left(2\alpha-1\right)+1}{2\left(1-\alpha\right)},\label{genHn}
\end{align}
where $H^{(n)}$ denotes the $n^{\textrm{th}}$ derivatives of the Hubble parameter with respect to the cosmic time.
Finally, we solve equation (\ref{genH}) to get the dependence of $x$ with respect to the cosmic time\footnote{The solution of $x$ in the case of $\alpha=1/2$ is given by $x=\frac{2}{3A}\rho_{\textrm{d}0}^{\frac{1}{2}}\left\{\exp\left[\left(6\pi G\right)^{\frac{1}{2}}A(t-t_{0})\right]-1\right\}$. On the other hand, if $\alpha=1$, the solution is given by $x=-\frac{2}{3A}\ln\left[\left(6\pi G \rho_{\textrm{d}0}\right)^{\frac{1}{2}}A(t_{s}-t)\right]$, where $t_{s}$ is the time of singularity.}
\begin{eqnarray}\label{genx}
x=\frac{1}{3A\left(1-\alpha\right)}\left\{\left[\left(6\pi G\right)^{\frac{1}{2}}A\left(1-2\alpha\right)(t-t_{0})\right.\right.\nonumber \\
\left.\left.+\rho_{\textrm{d}0}^{\frac{1-2\alpha}{2}}\right]^\frac{2\left(1-\alpha\right)}{1-2\alpha}-\rho_{\textrm{d}0}^{\left(1-\alpha\right)}\right\}.\
\end{eqnarray}
We next analyse the phenomenology of the model (\ref{genEoS}) described by different values of $\alpha$:
\begin{itemize}
\item If $\alpha\leq0$, the Hubble parameter diverges at large scales and at an infinite cosmic time but its cosmic time derivative vanishes at that point. We could identify this event as a LSBR wherein the first cosmic time derivative of the Hubble parameter vanishes at an infinite cosmic time instead of being a non-vanishing finite constant. This event is even smoother than the usual LSBR event ($\alpha=0$). We could consider the limit 
$\alpha\rightarrow-\infty$,  in which we recover the $\Lambda$CDM paradigm, (it is equivalent to choosing $A\rightarrow 0$). In that case the Hubble parameter stands constant at large scales and the Universe approaches asymptotically a de Sitter (dS) space time. 

\item If $0< \alpha\leq1/2$, first, we deduce that $H$ and $\dot{H}$ diverge at an infinite cosmic time and an infinite scale factor, therefore, this event corresponds to a LR. Second, we should  distinguish between two different cases: i) if $\alpha=\alpha_{n}\equiv(n-1)/2n$, then the $n^{\textrm{th}}$  cosmic time derivative of the Hubble parameter is finite while the lower order derivatives blow up at an infinite cosmic time and an infinite scale factor. ii) if $\alpha_{n-1}<\alpha<\alpha_{n}$, then the $n^{\textrm{th}}$ derivative of $H$ vanishes while the lower order derivatives blow up at an infinite cosmic time. Hence, we deduce that between the LSBR abrupt event and  the BR singularity there is a range of different LR abrupt events that could be characterised by the order in which the cosmic time derivative of Hubble parameter is finite (or vanishes). Therefore, we could coin these types of events as  ``$n^{\textrm{th}}$ order LR'' abrupt event when $\alpha=\alpha_{n}$ and ``$n^{\textrm{th}}$ zero order LR'' abrupt event when $\alpha_{n-1}<\alpha<\alpha_{n}$.  For example, in the ``second order LR'' abrupt event ($n=2,\alpha_{2}=1/4$), $H$ and $\dot{H}$ blow up at an infinite cosmic time and an infinite scale factor but the second order derivative, $\ddot{H}$, stands finite. We could move towards higher orders, where for $n\rightarrow\infty$ we get $\alpha=1/2$, that is,  in the strongest version of the LR abrupt event, the Hubble parameter and all its successive cosmic time derivatives blow up at an infinite cosmic time and an infinite scale factor. We could  name this event as the ``infinite order LR''  abrupt event\footnote{We have omitted the ``first order LR'' abrupt event ($n=1,\alpha_{1}=0$) since  actually it coincides with a LSBR. Therefore, for a consistent notation, we exclude the $n=1$ choice from the set.}.

\item If $1/2<\alpha\leq1$, the scale factor, the Hubble parameter and its cosmic time derivatives blow up at a finite cosmic time. Therefore, we identify this event with a BR. 

\item If $1<\alpha$, the Hubble parameter and its cosmic time derivatives blow up not only at a finite cosmic time but also at a finite scale factor as well. Therefore, we identify this event with a Big Freeze (BF).
\end{itemize}

In summary, when $\alpha\rightarrow-\infty$ we recover the cosmological constant model and the Universe is asymptotically dS. When 
$\alpha\leq0$, the Universe faces a LSBR abrupt event. If $0< \alpha\leq1/2$, the Universe faces a LR abrupt event. If $1/2<\alpha\leq1$, the induced singularity is a BR while for $1<\alpha$ is a BF. We sumarise the discussed abrupt events and singularities in table~\ref{sintab}.

\begin{table}[h!]
\begin{center}
\noindent\begin{minipage}{1\columnwidth}%
\begin{center}
\begin{tabular}{|>{\centering}m{2.cm}|>{\centering}m{0.7cm}>{\centering}m{0.7cm}>{\centering}m{0.7cm}>{\centering}m{0.7cm}>{\centering}p{0.7cm}|>{\centering}p{2.4cm}|}
\hline 
Event & $t$  & $a$  & $H$  & $\dot{H}$  & $H^{(n)}$ & $\alpha$
\tabularnewline
\hline 
dS & $\infty$ & $\infty$ & $H_{\Lambda}$ & $0$ & $0$ & $\alpha\rightarrow-\infty$ \tabularnewline
\hline 
LSBR$^{\star}$ & $\infty$ & $\infty$ & $\infty$ & $0$ & $0$ & $\alpha<0$ 
 \tabularnewline
\hline 
LSBR$^{\phantom{a}}$ & $\infty$ & $\infty$ & $\infty$ & $\dot{H}_{ls,1}$ & $0$ & $\alpha=0$  
\tabularnewline
\hline 
 $n^{\textrm{th}}$ zero order LR & $\infty$ & $\infty$ & $\infty$ & $\infty$ & $0$ & $\alpha_{n-1}<\alpha<\alpha_{n}$\tabularnewline
\hline 
 $n^{\textrm{th}}$ order LR & $\infty$ & $\infty$ & $\infty$ & $\infty$ &$H^{(n)}_{ls,n}$ & $\alpha=\alpha_{n}$  \tabularnewline
\hline 
infinite order LR & $\infty$ & $\infty$ & $\infty$ & $\infty$ & $\infty$ & $\alpha=1/2$ \tabularnewline
\hline 
BR  & $t_{br}$ & $\infty$ & $\infty$ & $\infty$ & $\infty$  & $1/2<\alpha\leq1$\tabularnewline
\hline
BF  & $t_{bf}$ & $a_{bf}$ & $\infty$ & $\infty$ & $\infty$ & $1<\alpha$\tabularnewline
\hline  
\end{tabular}
\caption{\label{sintab}%
This table shows some of the features of the main cosmological events achieved with a EoS as given in Eq. (\ref{genEoS}). The first column refers to the cosmological  event, the second column resumes the set of diverging cosmological parameters while the third column shows the value of $\alpha$ for which the event occurs, where  $\alpha_{n}=(n-1)/2n$ for $1<n$.
}
\par\end{center}
\begin{center}
\par\end{center}%
\end{minipage}
\par\end{center}
\end{table}
We conclude that for $1/2<\alpha$, the singularity occurs at a finite cosmic time while for $1<\alpha$ it also occurs at a finite scale factor. Note that asymptotically the EoS parameter goes as $w_{\textrm{d}}\rightarrow-1$ as far as $\alpha<1$, it is constant, $w=-(1+A)$, when $\alpha=1$ and diverges for $1<\alpha$. On the other hand, we could  identify the limit $\alpha\rightarrow-\infty$ with the  $\Lambda$CDM paradigm. However, we emphasise that while in a LSBR abrupt event all the bound structures are unavoidably destroyed, this is not the case in a Universe dominated by a cosmological constant. The specific set of values of $\alpha$  such that the acceleration is strong enough to induce the destruction of the bound structures will involve further calculations (see, for example, section 4 in \cite{Bouhmadi-Lopez:2014cca}). At least, for $0\leq\alpha$ we expect an effective destruction of the structures to occur at a finite cosmic time \cite{Bouhmadi-Lopez:2014cca}. 

\

Finally, we trust that the values of $\alpha$ chosen to describe the three  phantom DE models addressed in the present work are the most representative in their own species. First, $\alpha=0$ represents the smoother model inducing a LS where we know that the bound structures are ripped apart. Second, $\alpha=1/2$ is the particular value that induces the strongest version of a LR. Third, $\alpha=1$ is the largest value wherein a BR occurs. By the way, $\alpha=1/2$  and $\alpha=1$ are special cases which need a careful analysis. We will not consider a model inducing a BF as it leads to a singularity at a finite scale factor unlike the BR, LR,  and LSBR.  We will regard $\Lambda$CDM model as a guideline to compare our results. Nevertheless, we emphasise that an appropriate fitting of an interacting model should constrain the parameters $A$, $\alpha$ and $\lambda$ simultaneously rather than  constraining $A$ and $\lambda$ with a previously chosen $\alpha$, as we do in the present work. The motivation of focusing on these representative models consists on understanding how the interaction affects  the fate of the Universe in each case. As we will show in section~\ref{induced_events}, once the interaction between DE and DM is switched on all the abrupt events turn into a BR singularity. We emphasise that the analysis followed in this section only considers phantom models ($0<A$). Standard models, ($A<0$), would require a further analysis that will be done elsewhere.

\section{Interacting DM-DE}\label{sec2}
In this section we describe three phantom DE models under study. We solve the continuity equation to get  the DE and CDM densities for each model. We also show their confidence contours of the MCMC analysis using the data corresponding to \cite{Bouali:2019whr}.

\

Using the conservation equation (\ref{coupled}) and the generalised EoS (\ref{genEoS}), the energy density of DE and CDM can be written as 
\begin{equation}\label{rhodIgen}
\rho_{\textrm{d}}=\left[\left(\frac{3A}{\lambda}+\rho_{\textrm{d}0}^{1-\alpha}\right)e^{\lambda\left(1-\alpha\right)x}-\frac{3A}{\lambda}\right]^\frac{1}{1-\alpha},
\end{equation}
\begin{align}\label{rhomIgen}
\rho_{\textrm{m}}=&D_{0}e^{-3x}-\frac{\lambda}{3}\left(-\frac{3A}{\lambda}\right)^{\frac{1}{1-\alpha}}{}_{2}F_{1}\left[\beta,\gamma,\delta,f(x)\right],
\end{align}
%
%
where $D_{0}$ is a constant such that $\rho_{\textrm{m}}(0)=\rho_{\textrm{m}0}$, where the subscript $0$ denotes present values and  $_{2}F_{1}\left[\beta,\gamma; \delta; f(x)\right]$ is the Hypergeometric function. In addition,  the parameters $\beta,\gamma,\delta$ and the argument $f(x)$ read \footnote{We use the expression in the integral form as given by Eq. 15.3.1 in \cite{Abramowitz}. 
}
\begin{align}\label{paramI}
\beta&\equiv\frac{1}{\alpha-1},\\
\gamma&\equiv\frac{3}{\lambda\left(1-\alpha\right)},\\
\delta&\equiv 1+\frac{3}{\lambda\left(1-\alpha\right)},\\
f(x)&\equiv\left(1+\frac{\lambda}{3A}\rho_{\textrm{d}0}^{1-\alpha}\right)e^{\lambda\left(1-\alpha\right)x}.
\end{align}
We next show the results for the special cases $\alpha=1$ (model A) and 
$\alpha=1/2$ (model B), and we analyse as well the particular value $\alpha=0$ (model C).

\subsection{MODEL A: $\alpha=1$}\label{modelA}
The model A of the present work coincides with the widely known  $w_{\textrm{d}}$CDM model. Without interaction between DE and CDM its EoS reads as follows:
\begin{equation}\label{BRw}
p_{\textrm{d}}=w_{\textrm{d}}\rho_{\textrm{d}}.
\end{equation}
This model induces a BR singularity. When considering interaction, we use the conservation equation (\ref{coupled}) and the EoS (\ref{BRw}) to get the expression with respect to the scale factor of both energy densities of DE and DM,
\begin{equation}\label{BRrhod}
\rho_{\textrm{d}}(a)=\rho_{\textrm{d0}}a^{ (\lambda-3(1+w_{\textrm{d}}))} ,
\end{equation}
\hspace{-1cm}
\begin{equation}\label{BRrhom}
\rho_{\textrm{m}}(a)=\left(\rho_{m0} +\frac{\lambda \rho_{\textrm{d0}}}{\lambda-3w_{\textrm{d}}}\right)a^{-3}   -\frac{\lambda \rho_{\textrm{d0}}}{\lambda-3w_{\textrm{d}}}a ^{(\lambda-3(1+w_{\textrm{d}}))}.
\end{equation}
We label this model as IBR. The $\Lambda$CDM  model  can be recovered by fixing $w_{\textrm{d}}=-1$  and  $\lambda=0$. The BR singularity continue to exist if the condition $  3(1+w_{\textrm{d}}) < \lambda$ is satisfied.
\begin{figure}[h]
\centering
\includegraphics[scale=0.492]{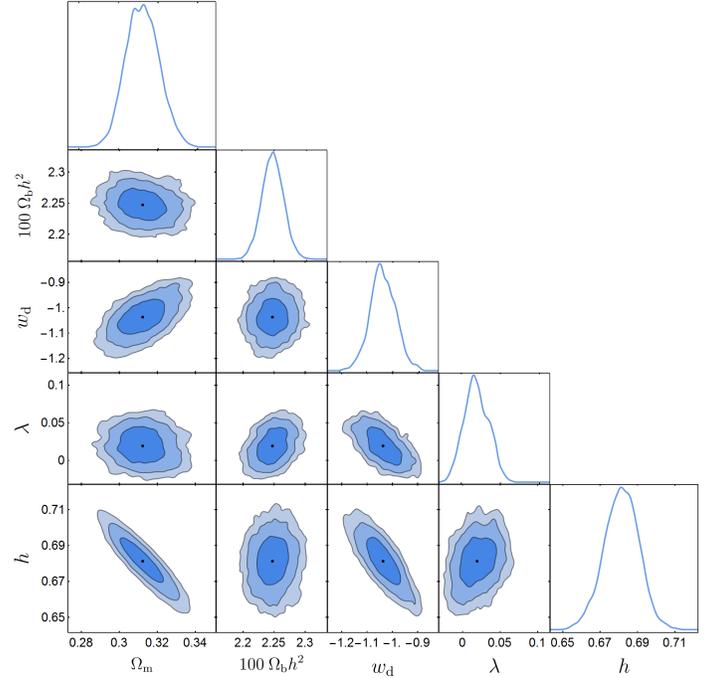}
\caption{ This figure corresponds to 1$\sigma$, 2$\sigma$ and 3$\sigma$ confidence contours obtained from SNIa+CMB+BAO+$\textrm{H}(z)$ data for the IBR model. }\label{contourBR}
\end{figure}

\subsection{MODEL B: $\alpha=1/2$}\label{modelB}

The second DE model under consideration is characterised by the following EoS \cite{Nojiri:2005sx,Stefancic:2004kb}
\begin{equation}\label{lr}
p_{\textrm{d}}=-\left( \rho_{\textrm{d}}+B\rho_{\textrm{d}}^{\frac{1}{2}}\right), 
\end{equation}
where $B$ is a positive  constant. Without interaction, it induces the abrupt event known as LR. Under interaction, the evolutions of DE and CDM  densities are obtained by solving the conservation equation Eq. (\ref{coupled}) and Eq. (\ref{lr})

\begin{equation}\label{LRrhod}
\rho_{\textrm{d}}(a)=\rho_{\textrm{d0}}\left[ a^{\lambda/2}+\frac{3}{\lambda}\sqrt{\frac{\Omega_{\textrm{lr}}}{\Omega_{\textrm{d0}}}}\left(  a^{\lambda/2}-1\right) \right]^2  ,
\end{equation}
\begin{widetext}
\begin{align}\label{LRrhom}
\rho_{\textrm{m}}(a)=&\Big\{   \rho_{m0}+\lambda \rho_{\textrm{d0}}\Big(\frac{1}{\lambda+3}-\frac{6}{(\lambda+3)(\lambda+6)}\sqrt{\frac{\Omega_{\textrm{lr}}}{\Omega_{\textrm{d0}}}}
-\frac{3(6-\lambda)}{\lambda^2(\lambda+6)}\frac{\Omega_{\textrm{lr}}}{\Omega_{\textrm{d0}}}\Big)\Big\}    a^{-3}\\
-&\frac{\lambda}{\lambda+3} \rho_{\textrm{d0}}\Big(1+\frac{3}{\lambda}\sqrt{\frac{\Omega_{\textrm{lr}}}{\Omega_{\textrm{d0}}}}\Big)^2a^{\lambda}
+\frac{3}{\lambda+6} \rho_{\textrm{d0}}\sqrt{\frac{\Omega_{\textrm{lr}}}{\Omega_{\textrm{d0}}}}\Big(1+\frac{3}{\lambda}\sqrt{\frac{\Omega_{\textrm{lr}}}{\Omega_{\textrm{d0}}}}\Big)^2a^{\lambda/2}-\frac{3}{\lambda} \rho_{\textrm{d0}}{\frac{\Omega_{\textrm{lr}}}{\Omega_{\textrm{d0}}}}\nonumber ,
\end{align}
\end{widetext}

where we have defined the dimensionless parameters $\Omega_{\textrm{lr}}$ and $\Omega_{\textrm{d0}}$ as
\begin{equation}
	\Omega_{\textrm{lr}}=\frac{8\pi G}{3H^2_0}B^2, \qquad \Omega_{\textrm{d0}}=\frac{8\pi G}{3H_0^2}\rho_{\textrm{d0}}.
\end{equation}

\begin{figure}[h]
\centering
\includegraphics[scale=0.5]{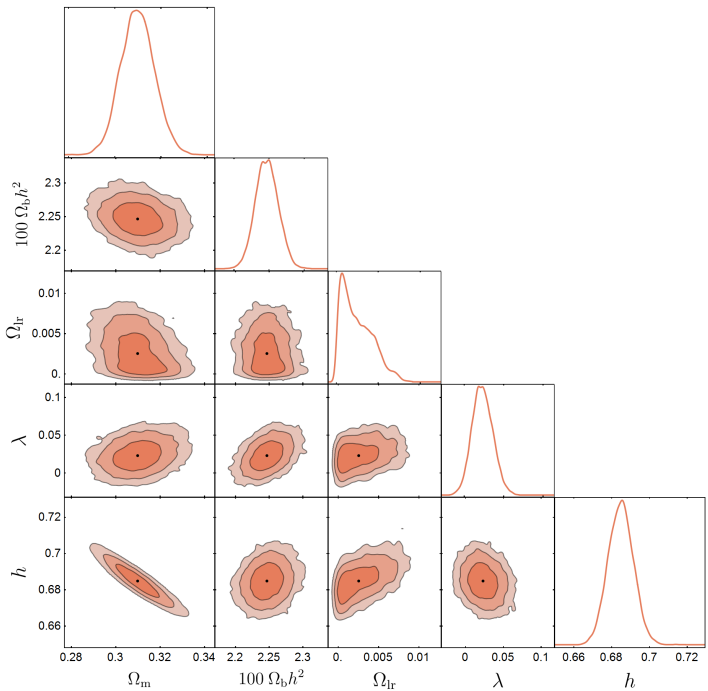} 
\caption{  This figure corresponds to 1$\sigma$, 2$\sigma$ and 3$\sigma$ confidence contours obtained from SNIa+CMB+BAO+$\textrm{H}(z)$ data for the ILR model. }\label{contourLR}
\end{figure}

\subsection{MODEL C: $\alpha=0$}\label{modelC}

The third model  deviates from the $\Lambda$CDM's EoS by a positive constant parameter $C/3$ as \cite{Bouhmadi-Lopez:2014cca}
\begin{equation}\label{lsbrrho}
p_{\textrm{d}}=-\left( \rho_{\textrm{d}}+\frac{C}{3}\right).
\end{equation}
Without interaction, this model induces the abrupt event known as LSBR. When considering interaction, we use the conservation equation (\ref{coupled}) and Eq.~(\ref{lsbrrho}) to obtain the evolution of DE and CDM densities

\begin{equation}\label{LSBRrhod}
\rho_{\textrm{d}}(a)=\rho_{\textrm{d0}}\left(\Big(1+\frac{\Omega_{\textrm{lsbr}}}{\lambda\Omega_{\textrm{d0}}}\Big)a^{\lambda}-\frac{\Omega_{\textrm{lsbr}}}{\lambda\Omega_{\textrm{d0}}}\right),
\end{equation}

\begin{eqnarray}\label{LSBRrhom}
\rho_{\textrm{m}}(a)&=&\left\{\rho_{m0}+ \frac{\lambda \rho_{\textrm{d0}}}{\lambda+3}\left(1-\frac{\Omega_{\textrm{lsbr}}}{3 \Omega_{\textrm{d0}}}\right)\right\}a^{-3}\nonumber\\
&-&\frac{\lambda \rho_{\textrm{d0}}}{\lambda+3}\Big(1+\frac{\Omega_{\textrm{lsbr}}}{\lambda\Omega_{\textrm{d0}}}\Big)a^{\lambda}
+\rho_{\textrm{d0}} \frac{\Omega_{\textrm{lsbr}}}{3 \Omega_{\textrm{d0}}},
\end{eqnarray}

where the dimensionless quantity $\Omega_{\textrm{lsbr}}$ is related to the positive parameter $C$ as $\Omega_{\textrm{lsbr}}\equiv (8\pi G/3H_{0}^{2})C$. 


\section{The cosmological events induced by models under interaction}\label{induced_events}

In a scenario where matter and DE are separately conserved, the models introduced in \ref{modelA}, \ref{modelB} and \ref{modelC} induce the cosmological events known as BR, LR and LSBR, respectively. Nevertheless, the end of the Universe would not be the same when  an  interaction between matter and DE is considered. The cosmological singularities are often classified according to the divergence of some of the cosmological magnitudes, as for example, the Hubble parameter and its cosmic time derivatives \cite{Bouhmadi-Lopez:2019zvz}. In addition, we refer to a singularity when it occurs at a finite cosmic time (as it is the case of BR) while we coin it as an abrupt event when it occurs at an infinite cosmic time (as it is the case of LR and LSBR) \cite{Albarran:2016mdu}.

\

In this section, we analyse the asymptotic behaviour of the first cosmic time derivative of the Hubble parameter, $\dot{H}$, in order to appropriately classify the cosmic event induced by each model under DE  and CDM interaction. 
Before proceeding with the calculations some comments are in order. From a  first glance, the expressions for the energy density of CDM  and DE (see Eqs. \eqref{BRrhod} and \eqref{BRrhom} for the model IBR, Eqs.\eqref{BRrhod} and \eqref{LRrhom} for the model ILR and  Eqs. \eqref{LSBRrhod} and \eqref{LSBRrhom} for the model ILSBR) show a power law with respect to the scale factor. If the highest power of the scale factor is positive, which is indeed our case in all the interaction models considered here, the scale factor diverges at a finite cosmic time. Therefore, we are dealing with true curvature singularities rather than abrupt events. 

\

In order to get the first cosmic time derivative of the Hubble parameter, we first derive the Friedmann equation with respect to the cosmic time, 

\begin{equation}\label{friedmann_der}
2H\dot{H}=\frac{\kappa}{3}\left(\dot{\rho}_{\textrm{m}}+\dot{\rho}_{\textrm{d}}\right),
\end{equation}

where $\kappa=8\pi G$. Then, we replace the conservation equation \eqref{coupled} to get
\begin{equation}\label{dotH}
\dot{H}=-\frac{\kappa}{2}\left(\rho_{\textrm{m}}+\rho_{\textrm{d}}+p_{\textrm{d}}\right).
\end{equation}
%
Finally, it just  remains  to replace the equations \eqref{BRw}-\eqref{LSBRrhom} above. We next show the asymptotic behaviour of $\dot{H}$ for the models with interaction considered in the present work.

\begin{itemize}
\item Model IBR
\begin{equation}\label{HdotBR}
\dot{H}\approx-\frac{3}{2}H_{0}^{2}\Omega_{\textrm{d}0}\left(\frac{\lambda}{\lambda-3w_{\textrm{d}}}+1+w_{\textrm{d}}\right)a^{\lambda-3\left(1+w_{\textrm{d}}\right)},
\end{equation}
\item Model ILR
\begin{equation}\label{HdotBR}
\dot{H}\approx\frac{3}{2}H_{0}^{2}\Omega_{\textrm{d}0}\frac{\lambda}{\lambda+3}\left(1+\frac{3}{\lambda}\sqrt{\frac{\Omega_{\textrm{lr}}}{\Omega_{\textrm{d}0}}}\right)^{2}a^{\lambda},
\end{equation}
\item Model ILSBR
\begin{equation}\label{HdotBR}
\dot{H}\approx\frac{3}{2}H_{0}^{2}\frac{\Omega_{\textrm{d}0}}{1+\lambda}\left(\lambda+\frac{\Omega_{\textrm{lsbr}}}{\Omega_{\textrm{d}0}}\right)a^{\lambda}.
\end{equation}
\end{itemize}

We remind that the parameter $\lambda$ is positive and 
$\kappa\rho_{\textrm{i}0}\equiv 3\Omega_{\textrm{i}0}H_{0}^{2}$ (i=r,d,m)  following the definitions given in  the expression \eqref{defOmega}. As can be seen, for all the interacting scenarios considered in the present work 
$\dot{H}$ diverges at very large scale factors. Consequently, it follows that the induced singularities are like BR type. Surprisingly,  phantom DE models in absence of interaction  induce a  smoother  versions of the BR  (LR and LSBR).  However, they  lead to a true BR singularity when switching on the interaction with matter.


\begin{figure}[h]
\centering
\includegraphics[scale=0.5]{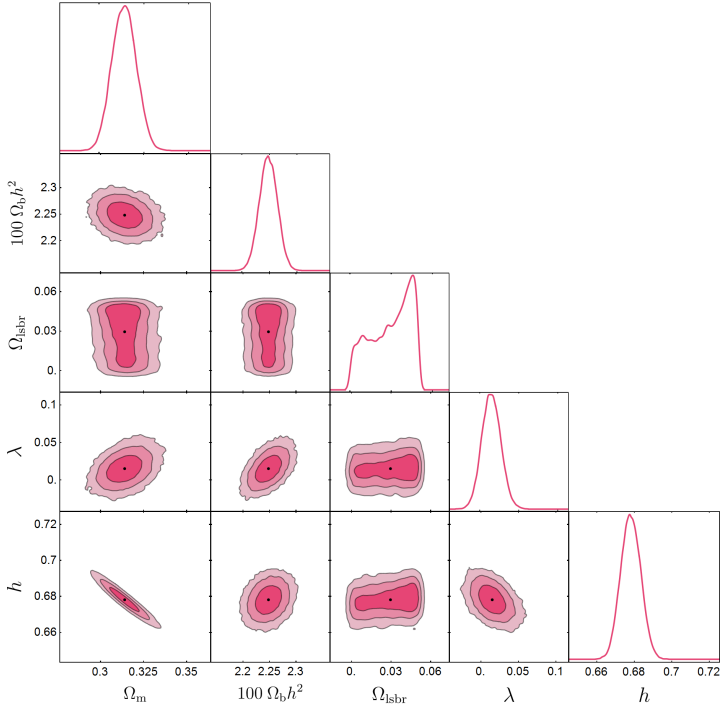} 
\caption{ This figure corresponds to 1$\sigma$, 2$\sigma$ and 3$\sigma$ confidence contours obtained from SNIa+CMB+BAO+$\textrm{H}(z)$ data for the ILSBR model.}\label{contourLSBR}
\end{figure}
\begin{figure}[h]\label{energy_tranfer}
\centering
\includegraphics[scale=0.55]{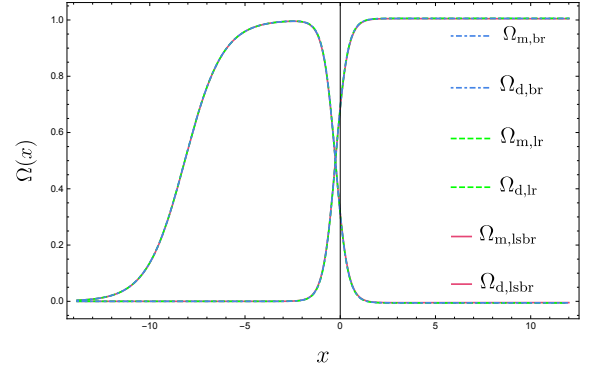} 
\caption{ This figure shows the energy tranfer between CDM and DE densities. The plot shows the decaying process of CDM density into DE density. The three models IBR, ILR and ILSBR show almost a perfect overlap. The black vertical line denotes the present. We remind that $x=\ln(a)$ and $a=1$; i.e. $x=0$, at present.}\label{energy_tranfer}
\end{figure}
\section{RESULTS}\label{sec3}
In order to discuss  the results  carried out by this analysis, we present as a first step  the priors used to constrain the cosmological parameters. For all models, we have allowed  matter density, the product of baryons density times the  square of the Hubble rate, the Hubble rate and the strength of  the  interaction to vary as $\Omega_{\textrm{m}}\in \left\lbrace 0.001,0.5\right\rbrace $,  $\Omega_{\textrm{b}} h^2 \in \left\lbrace 0.001,0.08\right\rbrace $, $ h \in \left\lbrace 0.4,1\right\rbrace $ and $\lambda \in \left\lbrace -1,1\right\rbrace $, respectively \footnote{The same priors were used in the non interacting case, please see \cite{Bouali:2019whr}.}. Besides the dimensionless parameters $\Omega_{lr}$ and $\Omega_{lsbr}$ vary in $\left\lbrace 0,1.7\times 10^{-3}\right\rbrace $ and $\left\lbrace 0, 5.025\times 10^{-2}\right\rbrace$, respectively. The choice of the dimensionless parameters $\Omega_{lr}$ and $\Omega_{lsbr}$ is done to compare our interacting models with the non interacting ones \cite{Bouali:2019whr}. Indeed, in the non interacting case this choice is mandatory to avoid negative values of the  DE density in the past, in particular, from the beginning of our numerical integration\footnote{Notice that the value $a_{\textrm{ini}}$ could be arbitrary fixed, however, it should be small enough to be in a completely radiation dominated epoch and large enough to stand outside the inflationary era.}, i.e. $a_{\textrm{ini}}=10^{-6}$.  From table~\ref{tab1},  we observe that for I$\Lambda$CDM model the amount of matter i.e. $\Omega_{\textrm{m}}$ increases, while the Hubble rate $h$ decreases slightly when compared to $\Lambda$CDM. From the confidence contours in figures \ref{contourBR}, \ref{contourLR} and \ref{contourLSBR}, we remark that the couple ($\omega_{\textrm{d}}$,$\lambda$) presents a negative correlation, while  the couples ($\Omega_{\textrm{lr}}$,$\lambda$) and ($\Omega_{\textrm{lsbr}}$,$\lambda$) present a positive correlation. We notice as well that our results of $H_{0}$ are far away from the one given by local measurements, e.g by Riess et al \cite{Riess:2019cxk}. This mismatch between  the $H_0$ values has become very troublesome and dubbed as the $H_0$ tension problem. As an attempt to solve this issue,   physics beyond the standard model must be advocated, for example higher number of effective relativistic species and non-zero curvature can be helpful to smooth this problem (see the interesting review \cite{DiValentino:2021izs} for further information on this topic). \\ 

 In order to classify the phantom DE models studied in this paper, based on their statistical ability to fit observational points, we use the corrected Akaike Information Criterion ($AIC_c$) defined as \cite{AIC}
\begin{equation}\label{AIC}
AIC_c = -2\ln{\cal{L}}_{max}+2k+\frac{2k(k+1)}{N_{\textrm{d}}-k-1},
\end{equation}
where $k$ and $N_{\textrm{d}}$ denote the number of free parameters  and  the total number of  observational data, respectively. For a random Gaussian distributions chi squared takes the following form 
 $$\chi^2_{min}=-2\ln{\cal{L}}_{max}.$$
 
The $AIC_c$ is a powerful tool to compare models having a different number of free  parameters, as well as a different number of data. Indeed, the model corresponding to the minimal value of $AIC_c$  is the most favoured by the observational data and it is considered as the reference model ($\Lambda$CDM is often the case). Once the reference model is determined, we calculate $AIC_c$ difference of each model with respect to it, i.e. $\Delta AIC_c =  AIC_{c,model}- AIC_{c,min}$. In general,  models with $0 < \Delta \textrm{AIC}_c< 2$ have substantial support, those with $4 < \Delta \textrm{AIC}_c< 7$ have considerably less support, and  models with $\Delta \textrm{AIC}_c> 10$ have essentially no support, with respect to the reference model.

\begin{table*}
\begin{center}
\begin{tabular}{cccccc}
\hline   
                   \bf Model              &\bf Par                &\bf Best fit     &\bf Mean          \\
\hline
\hline
\multirow{2}{*} {$\Lambda$CDM}          &$\Omega_{\textrm{m}}$               &$0.312308_{-0.00607812}^{+0.00607812}$  &$0.312583_{-0.00607362}^{+0.00607362}$     \\ [0.1cm]                                                                                              
                                        &$h$                     &$0.678603_{-0.00447778}^{+0.00447778}$      &$0.678435	_{-0.00447417}^{+0.00447417}$     &\\[0.1cm]
                                        &$\Omega_b h^{2}$        &$0.0224102_{-0.000134672}^{+0.000134672} $  &$0.0224002_{-0.00013449}^{+0.00013449}$          &\\[0.1cm]                                         
\hline
\multirow{2}{*} {I$\Lambda$CDM}          &$\Omega_{\textrm{m}}$               &$0.314738_{-0.00663245}^{+0.00663245}$  &$0.315252_{-0.00663212}^{+0.00663212}$     \\ [0.1cm]                                                                                              
                                        &$h$                     &$0.676461_{-0.00504971}^{+0.00504971}$      &$0.675976_{-0.005048282}^{+0.005048282}$     &\\[0.1cm]
                                         &$\lambda$              &$0.00992076_{-0.0116578}^{+0.0116578}$                                         &$0.011897_{-0.011643}^{+0.011643}$        & \\[0.1cm]
                                        &$\Omega_b h^{2}$        &$0.0224698_{-0.000159949}^{+0.000159949} $  &$0.0224845_{-0.000159746}^{+0.000159746}$          &\\[0.1cm]                                         
\hline                                                                     
\multirow{3}{*} {IBR}                    &$\Omega_{\textrm{m}}$              &$0.310664_{-0.00830269}^{+0.00830269}$    &$0.31219_{-0.00828119}^{+0.00828119}$             \\ [0.1cm] 
                                        &$w_{\textrm{d}}$          &$-1.03248_{-0.0482072}^{+0.0482072}$     &$-1.03728_{-0.0482912}^{+0.0482912}$     &\\[0.1cm]
                                        &$\lambda$              &$0.0168888_{-0.0154053}^{+0.0154053}$                                         &$0.0193522_{-0.0155017}^{+0.0155017}$                                             & \\[0.1cm]
                                        &$h$                     &$0.682033_{-0.0092436}^{+0.0092436} $     &$0.681365_{-0.00922296}^{+0.00922296} $       &\\[0.1cm]
                                        &$\Omega_b h^{2}$        &$0.0224867_{-0.000166489}^{+0.000166489} $  &$0.0224764_{-0.000166426}^{+0.000166426} $          &\\[0.1cm]                                        
                                        
\hline               
 \multirow{3}{*} {ILR}                   &$\Omega_{\textrm{m}}$              &$0.310516_{-0.00726847}^{+0.00726847}$     &$0.309924_{- 0.00726091}^{+0.00726091}$   \\ [0.1cm] 
                                        &$\Omega_{\textrm{lr}}$           &$0.00105782_{-0.00182933}^{+0.00182933}$    &$0.00252318_{-0.0018261}^{+0.0018261}$ &\\[0.1cm]
                                        &$\lambda$               &$0.017768_{-0.0129577}^{+0.0129577}$                                                                       &$0.0231745_{-0.0129438}^{+0.0129438}$                                                                                &\\[0.1cm]
                                        &$h$                     &$0.682915_{-0.00661987}^{+0.00661987} $   &$0.684813_{-0.00661268}^{+0.00661268} $        &\\[0.1cm]
                                        &$\Omega_b h^{2}$        &$0.0224682_{-0.00016143}^{+0.00016143} $  &$0.0224685_{-0.000161631}^{+0.000161631} $          &\\[0.1cm]                                        
                                        
\hline  

\multirow{3}{*} {ILSBR}                  &$\Omega_{\textrm{m}}$              &$0.313552^{+0.00671554}_{-0.00671554}$  &$0.314018^{+0.00670695}_{-0.00670695}$   \\  
                                        &$\Omega_{lsbr}$         &$0.0458419_{-0.0144588}^{+0.0144588} $     &$0.0295684_{-0.0145073}^{+0.0145073} $ &          &\\[0.1cm]
                                        &$\lambda$               &$0.0154082_{-0.0120793}^{+0.0120793}$   &$0.0148823_{-0.0120577}^{+0.0120577}$                                                                                                             &\\[0.1cm]                   
                                        &$h$                     &$0.679071_{-0.00524713}^{+0.00524713} $      &$0.678149_{-0.00524195}^{+0.00524195} $     & \\[0.1cm]
                                        &$\Omega_b h^{2}$        &$0.0224517_{-0.000162054}^{+0.000162054}$    &$0.0224808_{-0.000161849}^{+0.000161849}$       &\\[0.1cm]                                        
                                        
\hline
\hline                                     
\end{tabular}

\caption{Summary of the best fit and the mean values of the cosmological parameters.  }\label{tab1}

\end{center}
\end{table*}

\begin{table}[h]\label{tab2}
\begin{center}
\begin{tabular}{cccccc}
\hline  
                   &\bf Model       &$\bf {{\chi}^2_{red}}$  &$\bf {{\chi}^2_{tot}}^{min}$   &\bf$ AIC_{c}$  &\bf$\Delta AIC_c$ \\
\hline
\hline
&$\Lambda$CDM     &0.981699  &1073.9795             &1080.0014    &0 \\ [0.1cm]                                                                                              
                                                                          
\hline
& I$\Lambda$CDM   &0.981800 &1073.1076              &1081.1443    &1.1429\\ [0.1cm]                                                                                                                                                                            
\hline                                                                     
& IBR             &0.982314  &1072.6870              &1082.7420    &2.7406 \\ [0.1cm] 
                                                                                 
\hline                
& ILR             &0.982278  &1072.6477             &1082.7027    &2.7013 \\ [0.1cm] 
                                                                                                                       
\hline  
&ILSBR           &0.982289  &1072.6576              &1082.7126   &2.7111 \\   [0.1cm]
                                                                                                                   
\hline                                    
\end{tabular}

\caption{Summary of the $ {\chi}^{2}_{min}$, $AIC$, and $\Delta AIC$.  }

\end{center}
\end{table}

\

In table \ref{tab1} , we present a summary of the MCMC analysis of our phantom DE models $\Lambda$CDM, I$\Lambda$CDM, IBR, ILR and ILSBR. The model corresponding to the smallest value of $ AIC_{c}$ is the most favoured by observations i.e. $\Lambda$CDM in this case. Thus, we take $\Lambda$CDM  as reference model by fixing its corresponding $\Delta AIC_{\Lambda\textrm{CDM}}$ to 0. Models are classified with respect to the reference model. By calculating the $\Delta AIC$ quantity for each models,  we can see that I$\Lambda$CDM  is the closest model to $\Lambda$CDM followed by ILR, ILSBR and IBR, respectively. This classification is in discordance with that of the case without interaction \cite{Bouali:2019whr}.
\begin{figure}[h]
\hspace{-0.5cm}
\includegraphics[scale=0.55]{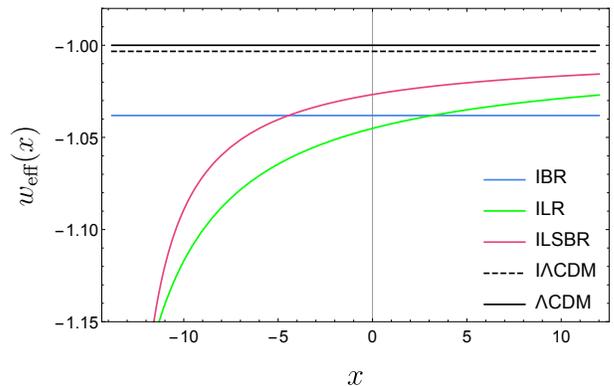} 
\caption{This figure shows the evolution of the effective EoS i.e. $w_{\textrm{eff}}=w_{\textrm{d}}-Q/3H\rho$ of $\Lambda$CDM, I$\Lambda$CDM, IBR, ILR and ILSBR models with respest to $x=\ln(a)$.  The black vertical line denotes the present. We remind that $x=\ln(a)$ and $a=1$; i.e. $x=0$, at present.}\label{EoS}
\end{figure}

We bring to the attention of the reader that in the case of a constant EoS i.e. $w_{\textrm{d}}=const$, one has to choose the coupling strength $\lambda$ and the EoS   carefully. In \cite{Valiviita:2008iv,Clemson:2011an,Gavela:2010tm},  the authors pointed out that for  $w_{\textrm{d}}$CDM model, the  interaction strength  $\lambda$ and the quantity (1+$w_{\textrm{d}}$) must have the same signs for $Q\propto \rho_d$.  A positive coupling is favoured by observations. Thus, our results are in agreement with \cite{Clemson:2011an,Gavela:2009cy,Gavela:2010tm,Salvatelli:2013wra}. In addition, models that are the object of this study have a phantom behaviour. This combination could give rise to some instabilities at the perturbative level. Even though, fitting observationally the cosmological perturbations is not the purpose of this paper, a perturbative analysis  is considered qualitatively in order to compare the models under study and determine which model is the most favoured by observations.
 
\section{PERTURBATIONS}\label{sec4}
In this section, we present a brief summary of the analysis of the linear cosmological perturbations of our interacting DE models. To this aim, we consider only scalar perturbations and the perturbed metric reads in the conformal time as
\begin{align}\label{petric_perturbation}
ds^{2}=a^{2}&\left\{-\left(1+2\Phi\right)d\eta^{2}\right. \nonumber \\ 
&\left.+\left[\left(1-2\Psi\right)\delta_{ij}\right]dx^{i}dx^{j}\right\},
\end{align}
where $\eta$ is the conformal time, i.e.  $d\eta=dt/a$. From now on, a prime will denote the derivative with respect to this conformal time. The physical quantities $\Phi$ and $\Psi$ are known as the Bardeen potentials. In the next step, we 
adopt the longitudinal (Newtonian) gauge, as can be seen by (\ref{petric_perturbation}) and we assume that none of the A-fluid introduces anisotropies at the linear perturbative level, i.e. the anisotropic stress $\pi_A=0$, which leads to the metric potentials equality $\Phi=\Psi$. Hence, the evolution equations of the dimensionless density perturbation $\delta_A=\delta\rho_A/\rho_A$ and  the velocity perturbation $\theta_A$ are given by \cite{Valiviita:2008iv} 
\begin{align}\label{delta}
&\delta^{\prime}_{A}+3\mathcal{H}(c^{2}_{sA}-w_{\textrm{A}})\delta_{A}+(1+w_{\textrm{A}})\theta_{A}\\
&+3\mathcal{H}[3\mathcal{H}(1+w_{\textrm{A}})(c^{2}_{sA}-c^{2}_{aA})]\frac{\theta_{A}}{k^{2}}-3(1+w_{\textrm{A}})\Psi^{\prime}\nonumber\\ 
&=\frac{aQ_{A}}{\rho_{A}}\left[\Psi-\delta_{A}+3\mathcal{H}(c^{2}_{sA}-c^{2}_{aA})\frac{\theta_{A}}{k^{2}} \right]+\frac{a}{\rho_{A}}\delta Q_{A}, \nonumber
\end{align}
\begin{align}\label{theta}
&\theta^{\prime}_{A}+\mathcal{H}(1-3c^{2}_{sA})\theta_{A}-\frac{c^{2}_{sA}}{\left( 1+w_{\textrm{A}}\right) }k^{2}\delta_{A}-k^{2}\Psi\\
&= \frac{aQ_{A}}{(1+w_{\textrm{A}})\rho_{A}}\left[ \theta-(1+c^{2}_{sA})\theta_{A}\right]-\frac{a}{(1+w_{\textrm{A}})\rho_{A}}k^{2}f_{A} \hspace{0.1cm},\nonumber
\end{align}
where $A$ stands for radiation, CDM or DE fluid.  The parameter $f_{A}$ describes the intrinsic momentum transfer of the fluid A  and  $c^{2}_{sA}$ is the sound speed in the A-fluid rest frame (r.f) defined as
\begin{equation}\label{defcs}
\left.c_{sA}^2\equiv\frac{\delta p_{A}}{\delta\rho_{A}}\right|_{\textrm{r.f}}.
\end{equation}

In the case of  barotropic fluids $p_{A}(\rho_{A})$, the A-fluid speed of sound, $\frac{\delta p_{A}}{\delta\rho_{A}}$, coincides with the A-fluid adiabatic speed of sound, $c_{a\textrm{A}}^2$, which  in the case of time varying EoS is defined as follows:

\begin{equation}\label{cacs}
c_{a\textrm{A}}^2 \equiv\frac{p'_{A}}{\rho'_{A}}=w_{\textrm{A}}+\frac{w_{\textrm{A}}^\prime \rho_{A}}{\rho_{\textrm{A}}^\prime}.
\end{equation}
That is, barotropic fluids are completely adiabatic. Nevertheless, when dealing with DE fluids, given that $c^{2}_{\textrm{ad}}<0$  instabilities  are induced  at the perturbation level \cite{Gordon:2004ez}. Therefore, in order to avoid those instabilities, it is convenient to consider a non adiabatic contribution on the DE pressure perturbation \cite{Bean:2003fb,Valiviita:2008iv}, 
\begin{eqnarray}\label{cacs2}
\delta p_{\textrm{d},non-adiabatic}=\left(c^{2}_{s\textrm{d}}-c^{2}_{a\textrm{d}}\right)\delta\rho_{\textrm{d}}\vert_{\textrm{r.f}} \ ,
\end{eqnarray}
where the  DE rest-frame speed of sound parameter, $c^{2}_{s\textrm{d}}$, it is a free parameter within the interval $[0,1]$ while the DE adiabatic speed of sound parameter, $c^{2}_{a\textrm{d}}$, could be time dependent. 

The continuity equations can be rewritten as
\begin{equation}\label{coupledbis}
\left\{
\begin{array}{l}
 \dot{\rho}_{\textrm{m}}+3H\rho_{\textrm{m}}=Q_{\textrm{m}}\\
\dot{\rho}_{\textrm{d}}+3H(\rho_{\textrm{d}}+p_{\textrm{d}})=Q_{\textrm{d}},
\end{array}
\right.
\end{equation} 
where
\begin{equation}
 aQ_{\textrm{d}}=-aQ_{\textrm{m}}=aQ=\lambda \mathcal{H}\rho_{\textrm{d}} \hspace{0.2cm} \textrm{and} \hspace{0.2cm}aQ_{\textrm{r}}=0.
\end{equation}
\par
The perturbation of the interaction term Q is given by
\begin{equation}
\delta Q=Q\left[\frac{\delta H}{H} +\delta_{\textrm{d}}\right],
\end{equation}
note that, when the interaction term is proportional to $H$, i.e. $Q\propto H$, we must deal with the perturbation Hubble rate in order to get  gauge invariant equations for the dark sector coupled models. 
 The expression of the perturbation of $H$ in the longitudinal gauge is written as \cite{Gavela:2010tm} 
\begin{equation}
\frac{\delta H}{H}\equiv \frac{1}{\mathcal{H}}\left[ \frac{\theta}{3}-\mathcal{H}\Psi-\Psi^{\prime}\right],
\end{equation}
where $\theta$ is the volume expansion rate given by the partial contributions of each A-fluid
\begin{equation}\label{pecvelocity}
\theta=\sum_{A}\frac{1+w_{A}}{1+w}\Omega_{A}\theta_{A},\qquad\textrm{where}\qquad w=\sum_{A}\Omega_{A}w_{A}.
\end{equation}
Finally, we can write the perturbation of the interaction term $Q=\lambda H \rho_{\textrm{d}}$ as

\begin{align}
a\delta Q_{\textrm{m}}&=-a\delta Q_{\textrm{d}}\nonumber \\
&=-\lambda \mathcal{H}\rho_{\textrm{d}}\left[ \frac{\theta}{3\mathcal{H}}-\Psi-\frac{\Psi^{\prime}}{\mathcal{H}}+\delta_{d} \right],
\end{align}

 In order to evaluate the growth rate, i.e. $f\sigma_8$, for the models under consideration we rewrite Eqs. (\ref{delta}) and (\ref{theta}) as follow \cite{Albarran:2016mdu} :

\begin{subequations}
\begin{align}\label{equ_sys}
	\left(\delta_{\textrm{r}}\right)_x &=\frac{4}{3}\left(\frac{k^2}{\mathcal{H}}v_{\textrm{r}}+3\Psi_x\right)
	,\\
	\left(v_{\textrm{r}}\right)_x &=-\frac{1}{\mathcal{H}}\left(\frac{1}{4}\delta_{\textrm{r}}+\Psi\right)
	,\\ 
	\left(\delta_{\textrm{m}}\right)_x &=\left(\frac{k^2}{\mathcal{H}}v_{\textrm{m}}+3\Psi_x\right)\nonumber \\
	&+\frac{\lambda\rho_{\textrm{d}}}{\rho_{\textrm{m}}}\left[ \frac{k^2v}{3\mathcal{H}}+\Psi_x-\delta_{\textrm{d}}+\delta_{\textrm{m}}\right] 
	,\\
	\left(v_{\textrm{m}}\right)_x &=-\left(v_{\textrm{m}}+\frac{\Psi}{\mathcal{H}}\right)
	,\\ 
	\left(\delta_{\textrm{d}}\right)_x &=\left( 1+w_{\textrm{d}}\right)\left\lbrace  \left[  \frac{k^2}{\mathcal{H}}+9\mathcal{H}(c_{s\textrm{d}}^2-c_{a\textrm{d}}^2)\right] v_{\textrm{d}}+3\Psi_x \right\rbrace  \nonumber\\
	&+3\left(w_{\textrm{d}}-c_{s\textrm{d}}^2\right)\delta_{\textrm{d}}\nonumber\\
	&-\lambda\left( \frac{k^2 v}{3\mathcal{H}}+\Psi_x+3\mathcal{H}(c_{s\textrm{d}}^2-c_{a\textrm{d}}^2)v_{\textrm{d}} \right)  
	,\\
	\left(v_{\textrm{d}}\right)_x &=-\frac{1}{\mathcal{H}} \left(\frac{c_{s\textrm{d}}^2}{1+w_{\textrm{d}}}\delta_{\textrm{d}}+\Psi\right)+(3c_{s\textrm{d}}^2-1) v_{\textrm{d}}\nonumber\\
	&+\frac{\lambda}{1+w_{\textrm{d}}}(v_{\textrm{m}}-(1+c_{s\textrm{d}}^2)v_{\textrm{d}}),\\ 
	 \nonumber
\end{align}
\end{subequations}
where $v$ is the total velocity potential related to $\theta$ as
\begin{equation}
v=-\frac{\theta }{k^2},
\end{equation}
$x=\ln a$ and  $(\;)_x$ is the derivative with respect to $x$. We have also used $(\;)^\prime=\mathcal{H}(\;)_x$ and $\theta_A=-k^2 v_A$. In this work, we assume $c^{2}_{s\textrm{d}}=1$ as it is the case in the scalar field representation \cite{Bean:2003fb,Valiviita:2008iv}. In addition, we have chosen the momentum transfer parallel to the four velocity of DM, i.e. $f_{A}=Q_{A}\left(v_{\textrm{m}}-v\right)$, in such a way that the momentum transfer vanishes on the rest frame of DM \cite{Li:2013bya}.

\subsection{Growth rate }
In this section, we analyse the growth rate theoretical predictions i.e. $f\sigma_{8}$ of each studied model. These predictions will be then confronted to the observational data. To this aim, two compilations are used, namely RSD-63 which contains 63 data points published since 2006 until 2018 and the RSD-22 data which is the most robust  22 data compilation  considered by the authors of \cite{Sagredo:2018ahx}   after analysing combinations of subsets in RSD-63 data points.\par

 The evolution of these theoretical curves  depend strongly  on the  best fit parameters deduced from the background analysis. The minimised $\chi^{2}_{f\sigma_8}$ of the growth rate is given by 
\begin{equation}
	\chi^2_{f\sigma_8}=\sum_{i=1}^{N}\left(\frac{f\sigma_{8,th}(z_{i})-f\sigma_8(z_i)}{\sigma_{f\sigma_8}}\right)^2,
\end{equation}
where $N=63$ $(22)$ for RSD-63 (RSD-22) data.


  \begin{table}
\begin{tabular}{ |p{2cm}|p{2cm}|p{2cm}|p{2cm}|}
\hline
 &\multicolumn{2}{|c|}{\bf $\chi^{2}_{f\sigma_{8}}$}\\
\hline
 \centering \bf Model  &  \centering $\bf RSD22$ &  \centering $\bf RSD63$ \tabularnewline
\hline
\centering {\bf $\Lambda$CDM}  & \centering {0.8181}  & \centering {0.7762}  \tabularnewline
\hline
\centering {\bf I$\Lambda$CDM}  & \centering {0.9993}  & \centering {0.9045}  \tabularnewline
\hline
 \centering  \bf IBR  & \centering 1.1439  & \centering 1.0236 \tabularnewline
 \hline
  \centering  \bf ILR  & \centering 1.1713 & \centering 1.0485 \tabularnewline
 \hline
\centering  \bf ILSBR  & \centering 1.1208  & \centering 1.00343 \tabularnewline
 \hline
\end{tabular}
   \caption{This table presents  the reduced $\chi^{2}_{f\sigma_{8}}$  values of  the perturbative analysis.
    These results are obtained after the confrontation of models with the growth dataset RSD-63 and RSD-22 shown in table~\ref{tab4}.}\label{tab3}
\end{table}

\newpage
Table \ref{tab3}, shows the reduced $\chi^{2}_{f\sigma_{8}}$ values for all models studied. The reduced $\chi^{2}$ is a powerful tool to test the goodness of  fit, it takes into account, the minimal value of $\chi^2$, the number of free parameters and the total number of data points. If we compare models, the best one is that whose $\chi^2_{red}$ value is closest to 1. $\Lambda$CDM  model presents an over fit for both RSD-22 and RSD-63 data. Indeed, this conclusion is due to the injection of the best fit parameters extracted from the background statistics in $\chi^{2}_{f\sigma_{8}}$. These results are viewed as a qualitative since they depend strongly on the background analysis. The interacting model I$\Lambda$CDM  shows a better fit of both RSD-22 and RSD-63 than $\Lambda$CDM. It seems also that the RSD-63 data fit the interacting models, IBR, ILR and ILBR better than RSD-22 data does.

  \begin{table*}\label{RSD}
  \begin{center}
\centering
\resizebox{0.7\textwidth}{!}{%
\begin{tabularx}{.85\textwidth}{ C{.3} C{.3} C{1} C{2.6} C{.4} C{1.2} }
\toprule
\toprule
	{\centering{\bf Index}}
	&\bm{$z$}
	& \bm{$f\sigma_8$}
	& {\bf Survey}
	& {\bf Ref.}
	& {\bf Year}
\\\midrule
    1
	&0.02
	& $0.314 \phantom{}\pm0.048\phantom{}$
	& 2MASS
	& \cite{Davis:2010sw}
	& $\textrm{13 November 2010}$
 \\
	2
	&0.02
	& $0.398 \phantom{}\pm0.065\phantom{}$
	& SNIa + IRAS
	& \cite{Turnbull:2011ty}
	&  $\textrm{20 October 2011}$
 \\
    3
	&0.02
	& $0.428 \phantom{}\pm0.046\phantom{}$
	& 6dF Galaxy Survey + SNIa
	& \cite{Huterer:2016uyq}
	& $\textrm{29 November 2016}$
 \\
    4
	&0.10
	& $0.370 \phantom{}\pm0.130\phantom{}$
	& SDSS-veloc
	& \cite{Feix:2015dla}
	&$\textrm{16 June 2015}$
 \\
    5
	&0.15
	& $0.490\phantom{0}\pm0.145\phantom{0}$
	& SDSS DR7 MGS
	& \cite{Howlett:2014opa}
	& $\textrm{30 January  2015}$
 \\
    6
	&0.17
	& $0.510\phantom{0}\pm0.060\phantom{0}$
	& 2dF Galaxy Redshift Survey
	& \cite{Percival:2004fs,Song:2008qt}
	&$\textrm{6 October 2009}$
 \\
    7
	&0.18
	& $0.360\phantom{}\pm0.090\phantom{}$
	& GAMA
	& \cite{Blake:2013nif}
	& $\textrm{22 September 2013}$
 \\
    8
	&0.25
	& $0.3512\pm0.0583$
	& SDSS II LRG
	& \cite{Samushia:2011cs}
	& $\textrm{9 December 2011}$
 \\
    9
	&0.32
	& $0.384\pm0.095$
	&  BOSS-LOWZ
	& \cite{Sanchez:2013tga}
	& $\textrm{17 December 2013}$
 \\
   10
	&0.37
	& $0.4602\pm0.0378$
	& SDSS II LRG
	& \cite{Samushia:2011cs}
	& $\textrm{9 December 2011}$
 \\
    11
	&0.38
	& $0.440\pm0.060$
	& GAMA
	& \cite{Blake:2013nif}
	& $\textrm{22 September 2013}$
 \\
    12 
	&0.44
	& $0.413\pm0.080$
	& WiggleZ Dark Energy Survey + Alcock-Paczynski distortion
	& \cite{Blake:2012pj}
	& $\textrm{12 June 2012}$
 \\
    13
	&0.59
	& $0.488\pm0.06\phantom{0}$
	& SDSS III BOSS DR12 CMASS
	& \cite{Chuang:2013wga}
	& $\textrm{8 June 2016}$
 \\
    14
	&0.60
	& $0.390\pm0.063$
	& WiggleZ Dark Energy Survey + Alcock-Paczynski distortion
	& \cite{Blake:2012pj}
	& $\textrm{12 June 2012}$
 \\
    15
	&0.60
	& $0.550\pm0.120$
	& Vipers PDR-2
	& \cite{Pezzotta:2016gbo}
	& $\textrm{16 Decembre 2016}$
 \\
    16
	&0.73
	& $0.437\pm0.072$
	& WiggleZ Dark Energy Survey + Alcock-Paczynski distortion
	& \cite{Blake:2012pj}
	& $\textrm{12 June 2012}$
 \\
    17
	&0.86
	& $0.400\pm0.110\phantom{}$
	& Vipers PDR-2
	& \cite{Pezzotta:2016gbo}
	& $\textrm{16 Decembre 2016}$
 \\
    18
	&0.978
	& $0.379\pm0.0.176$
	& SDSS-IV
	& \cite{Okada:2015vfa}
	& $\textrm{9 January 2018}$	
 \\
    19
	&1.23
	& $0.385\pm0.099$
	& SDSS-IV
	& \cite{Okada:2015vfa}
	& $\textrm{9 January 2018}$
 \\
    20
	&1.40
	& $0.482\pm0.116$
	& FastSound
	& \cite{Okada:2015vfa}
	& $\textrm{25 November 2015}$
 \\
    21
	&1.526
	& $0.342\pm0.070$
	& SDSS-IV
	& \cite{Okada:2015vfa}
	& $\textrm{9 January 2018}$
 \\
   22
	&1.944
	& $0.364\pm0.106$
	& SDSS-IV
	& \cite{Okada:2015vfa}
	& $\textrm{9 January 2018}$	
 \\
    23
	&0.35
	& $0.44\pm0.05$
	&SDSS-LRG
	& \cite{Okada:2015vfa}
	& $\textrm{30 October 2006}$	
 \\
    24
	&0.77
	& $0.490\pm0.18$
	& VVDS
	& \cite{Okada:2015vfa}
	& $\textrm{6 October 2009}$	
 \\
    25
	&0.25
	& $0.3665\pm0.0601$
	& SDSS-LRG-60
	& \cite{Okada:2015vfa}
	& $\textrm{9 December 2011}$	
 \\
    26
	&0.37
	& $0.4031\pm0.0586$
	& SDSS-LRG-60
	& \cite{Okada:2015vfa}
	& $\textrm{9 December 2011}$
 \\
    27
	&0.067
	& $0.423\pm0.055$
	&6dFGS
	& \cite{Okada:2015vfa}
	& $\textrm{4 July 2012}$	
 \\
    28
	&0.30
	& $0.407\pm0.055$
	& SDSS-BOSS
	& \cite{Okada:2015vfa}
	& $\textrm{11 August 2012}$
 \\
    29
	&0.40
	& $0.419\pm0.041$
	& SDSS-BOSS
	& \cite{Okada:2015vfa}
	& $\textrm{11 August 2012}$	
 \\
    30
	&0.50
	& $0.427\pm0.043$
	& SDSS-BOSS
	& \cite{Okada:2015vfa}
	&  $\textrm{11 August 2012}$	
 \\
    31
	&0.60
	& $0.433\pm0.067$
	& SDSS-BOSS
	& \cite{Okada:2015vfa}
	& $\textrm{11 August 2012}$	
 \\
    32
	&0.80
	& $0.47\pm0.08$
	& Vipers
	& \cite{Okada:2015vfa}
	& $\textrm{9 July 2013}$	
 \\
    33
	&0.35
	& $0.429\pm0.089$
	& SDSS-DR7-LRG
	& \cite{Okada:2015vfa}
	&$\textrm{8 August 2013}$	
 \\
    34
	&0.32
	& $0.48\pm0.1$
	& SDSS DR10 and DR11
	& \cite{Okada:2015vfa}
	& $\textrm{17 December 2013}$	
 \\
    35
	&0.57
	& $0.417\pm0.045$
	& SDSS DR10 and DR11
	& \cite{Okada:2015vfa}
	&  $\textrm{17 December 2013}$	
 \\
    36
	&0.38
	& $0.497\pm0.045$
	& BOSS DR12
	& \cite{Okada:2015vfa}
	& $\textrm{11 July 2016}$	
 \\
    37
	&0.51
	& $0.458\pm0.038$
	& BOSS DR12
	& \cite{Okada:2015vfa}
	& $\textrm{11 July 2016}$
 \\
    38
	&0.61
	& $0.436\pm0.034$
	& BOSS DR12
	& \cite{Okada:2015vfa}
	& $\textrm{11 July 2016}$	
 \\
    39
	&0.38
	& $0.477\pm0.051$
	& BOSS DR12
	& \cite{Okada:2015vfa}
	& $\textrm{11 July 2016}$
 \\
    40
	&0.51
	& $0.453\pm0.05$
	& BOSS DR12
	& \cite{Okada:2015vfa}
	& $\textrm{11 July 2016}$	
 \\
    41
	&0.61
	& $0.410\pm0.044$
	& BOSS DR12
	& \cite{Okada:2015vfa}
	& $\textrm{11 July 2016}$	
 \\
    42
	&0.76
	& $0.440\pm0.040$
	& Vipers v7
	& \cite{Okada:2015vfa}
	& $\textrm{26 October 2016}$
 \\
    43
	&1.05
	& $0.280\pm0.080$
	& Vipers v7
	& \cite{Okada:2015vfa}
	& $\textrm{26 October 2016}$	
 \\
    44
	&0.32
	& $0.427\pm0.056$
	& BOSS-LOWZ
	& \cite{Okada:2015vfa}
	& $\textrm{26 October 2016}$	
 \\
    45
	&0.57
	& $0.426\pm0.029$
	& BOSS CMASS
	& \cite{Okada:2015vfa}
	& $\textrm{26 October 2016}$	
 \\
    46
	&0.727
	& $0.296\pm0.0765$
	& Vipers
	& \cite{Okada:2015vfa}
	& $\textrm{21 November 2016}$	
 \\
    47
	&0.6
	& $0.48\pm0.12$
	& Vipers
	& \cite{Okada:2015vfa}
	& $\textrm{16 December 2016}$	
 \\
    48
	&0.86
	& $0.48\pm0.1$
	& Vipers
	& \cite{Okada:2015vfa}
	& $\textrm{16 December 2016}$	
 \\
    49
	&0.1
	& $0.480\pm0.16$
	& SDSS DR13
	& \cite{Okada:2015vfa}
	& $\textrm{22 December 2016}$	
 \\
    50
	&0.001
	& $0.505\pm0.085$
	& 2MTF
	& \cite{Okada:2015vfa}
	&$\textrm{16 June 2017}$	
 \\
    51
	&0.85
	& $0.45\pm0.11$
	& Vipers PDR-2
	& \cite{Okada:2015vfa}
	& $\textrm{31 July 2017}$	
 \\
    52
	&0.31
	& $0.469\pm0.098$
	& BOSS DR12
	& \cite{Okada:2015vfa}
	& $\textrm{15 September 2017}$	
 \\
    53
	&0.36
	& $0.474\pm0.097$
	& BOSS DR12
	& \cite{Okada:2015vfa}
	& $\textrm{15 September 2017}$
 \\
    54
	&0.4
	& $0.473\pm0.086$
	& BOSS DR12
	& \cite{Okada:2015vfa}
	& $\textrm{15 September 2017}$	
 \\
    55
	&0.44
	& $0.481\pm0.076$
	& BOSS DR12
	& \cite{Okada:2015vfa}
	& $\textrm{15 September 2017}$	
 \\
    56
	&0.48
	& $0.482\pm0.067$
	& BOSS DR12
	& \cite{Okada:2015vfa}
	& $\textrm{15 September 2017}$	
 \\
    57
	&0.52
	& $0.488\pm0.065$
	& BOSS DR12
	& \cite{Okada:2015vfa}
	& $\textrm{15 September 2017}$	
 \\
    58
	&0.56
	& $0.482\pm0.067$
	& BOSS DR12
	& \cite{Okada:2015vfa}
	& $\textrm{15 September 2017}$	
 \\
    59
	&0.59
	& $0.481\pm0.066$
	& BOSS DR12
	& \cite{Okada:2015vfa}
	& $\textrm{15 September 2017}$	
 \\
    60
	&0.64
	& $0.486\pm0.07$
	& BOSS DR12
	& \cite{Okada:2015vfa}
	& $\textrm{15 September 2017}$	
 \\
    61
	&0.1
	& $0.376\pm0.038$
	& SDSS DR7
	& \cite{Okada:2015vfa}
	& $\textrm{12 December 2017}$	
 \\
    62
	&1.52
	& $0.42\pm0.076$
	& SDSS-IV
	& \cite{Okada:2015vfa}
	& $\textrm{8 Junuary 2018}$		
\\					
    63
	&1.52
	& $0.396\pm0.079$
	& SDSS-IV
	& \cite{Okada:2015vfa}
	& $\textrm{8 Junuary 2018}$		
\\											
\bottomrule
\bottomrule
\end{tabularx}
}
\caption{\label{tab4}%
This table shows a compilation of  $f\sigma_8$ redshift space distortion (RSD) data, published since 2006 until 2018.
}
\end{center}
\end{table*}

\section{CONCLUSION}\label{sec5}
  
The purpose of the current paper is to study  the behaviour of  three interacting phantom dark energy models, labeled as IBR, ILR and ILSBR, where the dark  components interact with each other, via a non gravitational interaction. We consider an interaction between DE and CDM densities to find out its impact on the cosmological parameters i.e. $\Omega_m$, $\Omega_d$ and  $h$,  as well as, to find out its impact on the persistence or the dissapearance of singularities and abrupt events. We have considered an interaction of the form $Q=\lambda H \rho_{\textrm{d}}$, because it does not give rise to a large scale instability at early times with $\lambda>0$ \cite{Li:2013bya}. We have studied models  at the background level as well as at the perturbative level. The background study consists  in  confronting the theoretical predictions of the models  with the joined observational data  corresponding to supernova from Pantheon, CMB from Planck 2018, BAO from (SDSS DR12, SDSS MGS, WiggleZ, 6dFGS, Lya, DES) and $H(z)$ by means of an MCMC approach, which allowed us to extracted the best fit parameters as well as their corresponding minimal chi squared values. In order to classify these models, we have used the $\textrm{AIC}_{c}$  tool. The results of the background statistical analysis reported that $\Lambda$CDM is  always the best model, followed by I$\Lambda$CDM, ILR, ILSBR,  and IBR. We remark that the interaction has changed the order of the preferred model in \cite{Bouali:2019whr}. As can be seen in table \ref{tab1}, observations favour a positive interaction  for all models i.e. CDM  decays into DE. We observe also that for IBR, ILR and ILSBR, the amount of the CDM energy density transformed to DE density is of order $10^{-2}$, while for I$\Lambda$CDM the amount of CDM density transformed into DE density is  of the order $10^{-3}$. Unfortunately, this process exacerbates the energy density transfer from  CDM into DE in the future, as shown in figure \ref{energy_tranfer}, by giving rise to a non physical results which breaks down the validity of this interaction  in the future.  From the MCMC approach, we have constrained the cosmological parameters of each phantom models. Table \ref{tab1} summarizes the best fit parameters of these models. By using the best fit parameters of the background analysis, we have solved numerically the perturbation equations system which allowed us to confront the predicted  $f\sigma_8$ of each model to observations i.e. RSD-22 and RSD-63. It was shown that the theoretical curves of IBR, ILR and ILSBR fit better the RSD-63 compilation. It seems that $\Lambda$CDM and I$\Lambda$CDM over fit both RDS-22 and RSD-63 compilations. Both RSD compilations reported that the classification of preferred models according to observations is as follows: $\Lambda$CDM, I$\Lambda$CDM, ILSBR, IBR and ILR model. In fact the perturbation results are not conclusive since they depend strongly on the background results. Consequently, our perturbative results  can be seen as  qualitative.   However, notice that our models are consistent with RSD compilations.

 Cosmological observations show that the singularity induced by the BR model as well as  the abrupt events  corresponding to the LR and the LSBR remain, even when the dark components interact with each other at least for the interaction  we have chosen.\\

\section{ ACKNOWLEDGMENTS}\label{sec6}

The work of M.B.L. is supported by the Basque Foundation of Science Ikerbasque. She also would like to acknowledge the support from the Basque government Grant No. IT956- 16 (Spain) and from the Grant
PID2020-114035GB-100 funded by MCIN/AEI/10.13039/501100011033 and by “ERDF A way of making Europe”.

\end{document}